\RequirePackage{fix-cm}
\documentclass[smallextended]{svjour3}       % onecolumn (second format)
\smartqed  % flush right qed marks, e.g. at end of proof
\usepackage{graphicx}
\usepackage{color}
\begin{document}
\title{Analyzing research performance: proposition of a 
new complementary index}
%\subtitle{Do you have a subtitle?\\ If so, write it here}

%\titlerunning{Short form of title}        % if too long for running head

\author{Shaon Sahoo }

%\authorrunning{Short form of author list} % if too long for running head

\institute{S. Sahoo \at
              CEA, INAC/SPSMS, F-38000 Grenoble, France \\
              \email{shaon.sahoo@cea.fr}           %  \\
%             \emph{Present address:} of F. Author  %  if needed
}

\date{Received: date / Accepted: date}
% The correct dates will be entered by the editor

\maketitle

\begin{abstract}
A researcher collaborating with many groups will normally have more papers (and
thus higher citations and $h$-index) than a researcher spending all his/her time 
working alone or in a small group. While analyzing an author's research merit, 
it is therefore not enough to consider only the collective impact of the 
published papers, it is also necessary to quantify his/her share in the impact. 
For this quantification, here I propose the $I$-index which is defined as an 
author's percentage share in the total citations that his/her papers have attracted.
It is argued that this $I$-index does not directly depend on the most of the
subjective issues like an author's influence, affiliation, seniority or career break.
A simple application of the Central Limit Theorem shows that, the scheme of
equidistribution of credit among the coauthors of a paper will give us the most
probable value of the $I$-index (with an associated small standard deviation which
decreases with increasing $h$-index).
I show that the total citations ($N_c$),
the $h$-index and the $I$-index are three independent parameters (within their bounds),
and together they give a comprehensive idea of an author's overall research performance.

% \PACS{PACS code1 \and PACS code2 \and more}
% \subclass{MSC code1 \and MSC code2 \and more}
\end{abstract}

\section{Introduction}
\label{intro}
At this age of increasing specialization, it has become almost impossible to go through
all the research works of an author and judge their merits. This inability necessitates
an objective analysis of an author's research output so that a wider population can
comprehend his/her research merit. This objective analysis is also very helpful in
comparing research outputs of different authors, and has become important tool for
the employers, policy makers and grand commissions.

How do we objectively and comprehensibly analyze an author's research merit? Clearly
for such an analysis many factors should be considered -the quality and quantity of
the research output, coauthors' contributions to a researcher's work, his/her ability
to do independent work, a researcher's efficiency in doing collaborative work,
his/her ability in working in different fields, etc. It is possible to carefully
define different parameters/metrics
to quantify each of the above aspects of an author's research performance.
To reflect on a particular aspect, if I may use physics terms, one is required to
extract the ``coarse-grained" information out of a huge amount of ``microscopic"
details associated with an author's publications and their impacts.

At this point it is important to realize that a single parameter or metric can
not give a full view of an author's scholarship or research merit.
As mentioned above, different parameters can be defined to judge different aspects 
of an author's scholarly output. For an efficient and objective description
of an individual's overall research performance, 
it is therefore crucial to recognize the most 
important aspects of research output and separately quantify them by carefully 
defined parameters. These parameters are expected to be
independent to avoid redundancy and it is also expected that they will have some
simple physical meaning such that they can be comprehended by the wider population.

In this work we identify three most important aspects of an author's research
output - (a) quantity, (b) quality and (c) author's own contribution in his/her
published works. In other way, these three aspects are the collective impact of 
the published papers,
author's productivity and author's share in the total impact of his/her works.
Clearly we need at least three independent parameters/metrics to
reliably quantify these three different aspects. What are the
three independent parameters which best serve this cause? I will argue in
this paper that the total number of citations ($N_c$), the $h$-index and the
newly defined $I$-index -these three parameters do the job satisfactorily.

Let me now briefly discuss why the division of
credit among the coauthors is so important, and $N_c$ and the $h$-index
are not enough to analyze an author's scholarly activity.
It is not uncommon for the senior and established researchers to
collaborate with many groups and publish a large numbers of papers per year.
These researchers will normally have higher citations ($N_c$) and the
$h$-index than those who are spending all their time working alone or
in a small group. It will be greatly
unfair for these lonely or small group workers if an author's research
performance is analyzed only by the parameters $N_c$ and the $h$-index.
It is therefore necessary to quantify a researcher's own role in his/her
success or in other words, how much the researcher could have achieved if
he/she had worked independently.
Here I propose the $I$-index (it can be interpreted as the
{\it Independence}-Index) to solve this problem. We will see in the next
section that, this index has a simple meaning
which will appeal to the wider population. It is defined in such a way
that its value will not directly depend on the most of the subjective
issues like an author's popularity/influence, affiliation, seniority and
career break/low activity (due to some severe medical condition, family
tragedy or importantly a female researcher's motherhood).
It is also argued in this paper that a simple scheme of equidistribution
of credit among the coauthors of a paper will not normally result in a
significant error in calculating the $I$-index.
We will see
that $N_c$, the $h$-index and the $I$-index are three independent
parameters (within their bounds), and together can give a comprehensive
idea about a researcher's overall performance
(see Sec. \ref{sec:3}).

There is an additional advantage in considering the $I$-index
while analyzing an individual's research output. The  parameters like $N_c$
and $h$-index can be unethically inflated in different ways. For example, a
number of
researchers working in several independent groups can decide that when a
group publishes a paper, it will give authorships to the members from other
groups even when they do not contribute. It is often complained that junior
authors are sometimes compelled to give authorships to senior non-contributing
researchers for sub-academic reasons. This unethical practice will be
discouraged if the $I$-index is considered
while analyzing an individual's research performance.

It is often stated that, even though it is very
important to quantify an author's own share in the total credit of his/her
published papers, but doing so may demoralize researchers to do true
collaborations which are imperative for the progress and betterment of science.
This crucial issue can be mostly resolved if we quantify three different
research aspects by three separate independent parameters. In this 
three-parameter framework of research analysis, researchers will be encouraged 
to do effective collaborations to improve their $N_c$ and $h$-index. At the 
same time they will be probably restrained from resorting to the unethical 
practices (mentioned above) if the $I$-index is also considered along 
with the other two indices. In this framework of analysis, authors' ranking  
can still be done according to their $h$ values (supplemented by $N_c$); the 
$I$-index can help resolve the ranking issues when multiple authors have 
close values of $h$-index (and $N_c$). In fact, researchers can be ranked in 
different ways depending on what importance is given to the $I$-index 
(for more discussions, see Sec. \ref{sec:3.1}).

In this work, {\it not} much importance is given to describe all three aspects
of research performance by a single parameter or metric. Here I may
emphasize that any attempt to do so would be gross due to serious loss of 
informations. The obscurity or ambiguity resulting from
the loss of informations may eventually lead in the error of judgement; as a
consequence, a group of scientists may get undue advantage while the deserving
candidates may be penalized. For example, though the $h$-index \cite{hirsch05}
somewhat successfully quantifies first two aspects of an
author's research output, the $\hbar$-index \cite{hirsch10}, which additionally
attempts to consider
coauthors' role, is not that successful. Besides loosing simple meaning and
calculation friendliness, the $\hbar$-index is known to be unfair towards junior
researchers and extra biased towards senior (having high $h$-index) researchers.
Three carefully-defined independent parameters would
provide us much better view (higher resolution) of a researcher's scholarly
activity than any single parameter can possibly do. 
With these facts in mind, one may also like to know what 
parameter we should use if for some practical reasons it is
needed to rank authors by a single parameter. For this purpose, 
in Sec. \ref{sec:3.1}, I define a normalized $h$-index (written as 
$\widetilde{h}$-index) which combines the effects/impacts of both $h$-index 
and $I$-index in a rational way. This $\widetilde{h}$-index is interpreted as 
the possible $h$-index of an author if he/she had worked alone. Subsequently 
I also propose $\widetilde{h}_T$-index which
additionally takes care of the seniority issue.

\section{$I$-index: definition and characteristics}
\label{sec:2}
Before I define and discuss the $I$-index, I will first briefly deliberate on two
main {\it assumptions} considered in this work:
\smallskip
\begin{enumerate}
 \item[(1)] {\it
The impact of a paper is solely determined by the number of citations it
received. This number of citations is the total credit to be distributed
among the coauthors of the paper.}
\smallskip
 \item[(2)] {\it For a multi-author paper, each author is indispensable
and effectively contributes equally if not mentioned otherwise.}
\end{enumerate}
\smallskip
While the first assumption is somewhat easy to comprehend, the second assumption
needs some discussions. 
I will argue and try to establish in this work that,
even though the assumption of equidistribution of credit may not be satisfactory
when applied to a single publication, it becomes quite a reasonable assumption
when applied to all the publications by an individual to determine his/her 
overall share in the total citations received by those publications.

Controversies and debates over credit distribution are not rare.
Despite the fact that it is crucial to distribute the credit among the
coauthors, the demarcation of contributions is a hopelessly difficult job.
Sometimes even for the coauthors it appears impossible to decide who contributed
what and what weight it carries. Sometimes an author's contribution may be small but
indispensable, without which the paper will not be complete and published. Sometimes
though a senior author's direct contribution to a paper is less but we have to
remember that he/she generally spends lot of time writing projects and bringing
fundings, without which there will be no research and no paper.
Any `logical' distribution of credit among the coauthors of a paper is highly
subjective and hence debatable. Different experts evaluating a multi-author paper
would give different credits to a particular author depending on how the evaluation
was done. 
This discussion clearly shows that, due to the inherent subjective 
nature of the analysis, we can not have a satisfactory deterministic model for 
quantifying an individual's share in the total credit of his/her published papers 
(the third aspect of research output, as mentioned before). 
If we define an index/metric to quantify this aspect of research output, and a large 
number of experts independently estimate the value of the index for an individual, 
then they will get different values for the index. Due to this inevitable randomness 
(or uncertainty) in the estimated value of the index, we need to develop a realistic 
statistical model to predict the most probable (or expectation) value of the index.
Here I define the $I$-index to quantify an individual's share
in the total credit of his/her works. I then discuss two relevant statistical models 
(two different statistical approaches), 
and show that, within the domain of their validity, the scheme of equidistribution of 
credit gives the most probable value of the $I$-index. Frequently in this paper the 
most probable value of the index is simply referred to as the $I$-index of an 
individual. It may be also mentioned here that the statistical arguments presented 
in this work is not generally applicable for a junior author with only a few papers.
\smallskip

{\underline{\it Definition}}: The $I$-index is an author's percentage share in
the total citations received by his/her published papers. If $c_i$ is the number 
of citations received by the $i$-th paper and $z_i$ is the author's
expected share of credit for the paper, then his/her $I$-index is given by:
\begin{eqnarray}
\label{iidxdf}
I = \frac{\sum_{i=1}^{N_p} z_i}{\sum_{i=1}^{N_p}c_i} \times 100 \%,
\end{eqnarray}
where $N_p$ is the total number of papers published by the author.
Now if $n_i$ is the number of authors contributed for the $i$-th
paper, then, assuming the equidistribution of credit among the coauthors, we have
$z_i = c_i/n_i$. Consequently, the author's $I$-index would be,
\begin{eqnarray}
\label{iindex}
I = \frac{\sum_{i=1}^{N_p} c_i/n_i}{N_c} \times 100 \%,
\end{eqnarray}
where $N_c = \sum_{i=1}^{N_p}c_i$.
\smallskip

In the following I will present two different statistical arguments to demonstrate 
the effectiveness of the equidistribution of credit scheme in calculating the $I$-index. 
After that I will discuss some of the main features or the characteristics of the index.
\smallskip

{\it Argument} (1): In short, here I will argue that  
the value given by Eq. \ref{iindex} is the most probable value 
or the expectation value of the $I$-index defined in Eq. \ref{iidxdf};
the statistical error in calculating the $I$-index using
Eq. \ref{iindex} is not normally significantly large.
 
Consider that a 
multi-author paper has $n$ coauthors and received $c$ citations. Let $z^j$ is 
the $j$-th author's expected share of credit for the paper; 
it is possible to express this quantity in the 
following form: $z^j = c/n + e^j$, where $e^j$ is the author's deviation of 
share from the average share of coauthors ($c/n$).
Since the total credit to be 
distributed among the $n$ authors is $c$, we must have $\sum_{j=1}^{n}z^j = c$.
This implies, $\sum_{j=1}^{n}e^j = 0$. Now using this relation and the fact 
that $z^j > 0$, we can get the strict mathematical bounds for $e^j$: 
$-\frac{c}{n} < e^j < \frac{(n-1)c}{n}$. In practice we expect the deviation 
$|e^j|$ to be small and within some fraction of the average share 
(i.e., $|e^j| < c/n$). The relation $\sum_{j=1}^{n}e^j = 0$ confirms that, the 
quantity $e^j$ would be positive for some authors 
and negative for others ($e^j$ can be zero, of course). Now which authors 
deserve to get positives values of $e^j$ and which authors should get negative 
values? While this can be hard to decide, it will not be  
unreasonable to assume here that, for an individual author with many published 
papers, his/her $e^j$ will be positive for some of his/her papers and 
negative for others. In other words, sometimes an individual researcher's 
contribution to a multi-author paper can be more than the coauthors' average 
contribution to the paper, while in some other occasions his/her contribution 
to a multi-author paper would be less than the average contribution. 
In the following I will use this statistical property of $e^j$ to 
calculate an individual's expected share in the total citations received
by his/her papers (the superscript index $j$ will be dropped since we will 
focus on a particular author). 

Let a researcher's expected share of credit for his/her 
$i$-th paper is $z_i = c_i/n_i + e_i$, where $e_i$ is a small 
number ($|e_i| < c_i/n_i$). While $c_i/n_i \ge 0$ for all papers, 
statistically the number $e_i$ would take positive values for some papers 
and negative values for others. When $n_i = 1$, we have $e_i = 0$, since for 
a single-author paper its sole author gets all the credit ($z_i = c_i$). 
Now when we calculate the researcher's total share in the collective credit 
of his/her papers by summing $z_i$ over all the published papers, we get 
$C_{share} = \sum_{i=1}^{N_p}c_i/n_i + E_r$, with 
$E_r = \sum_{i=1}^{N_p} e_i$. Since $e_i$ is a small quantity 
($|e_i| < c_i/n_i$) and statistically it takes both positive 
and negative values, we expect that $E_r$ will generally be a very 
small number when $N_p$ is large (i.e. $|E_r|\ll\sum_{i=1}^{N_p}c_i/n_i$).
Therefore, if we ignore $E_r$ and just take 
$C_{share} \approx \sum_{i=1}^{N_p}c_i/n_i$, 
then the resultant error would be normally less than what one might expect to 
get from this simple scheme of equidistribution of credit (in somewhat 
different context an argument similar in spirit can be found in 
Refs. \cite{pepe12,kurtz05}). While calculating the $I$-index, 
this resultant error ($E_r$) will then be further weakened due to the
presence of the large denominator factor ($N_c$) in the definition of 
the index (cf. Eq. \ref{iidxdf}). We note that the possible 
statistical error in calculating the $I$-index using Eq. \ref{iindex} is 
$\Delta = \frac{E_r}{N_c}\times100\%$. This error is expected to be negligible 
when $N_c$ becomes large.

Let us now try to get a rough estimation of $|\Delta|$ for 
an individual. 
First consider that the author has $l$ number of significant papers so 
that the total number of citations for these $l$
papers is much larger than the total number of citations for the rest of
the papers (i.e., $\sum_{i=1}^lc_i\gg\sum_{i=l+1}^{N_p}c_i$
when papers are arranged in the descending order of citation count).
The value of $l$ can be assumed to be the $h$-index of the
author. Furthermore consider that $\overline{c}$ and $\overline{n}$ are 
respectively the average number of citations and the average number of 
authors for those $l$ significant papers. As we discussed before, in
practice we expect $|e_i|$  to be some percentage of the corresponding average,
i.e., $|e_i| \sim \frac{x_i}{100} \times \frac{c_i}{n_i}$ where $x_i$ may take 
any value between, say, 0 and 20. This allows us to write  
$E_r = \sum_{i=1}^{N_p}e_i \sim 
\frac{\overline{c}}{100\overline{n}}\sum_{i=1}^{l}s_i x_i$. Here $s_i$ carries 
only the sign of $e_i$; if $e_i$ is positive (negative), 
then $s_i = +1$ ($s_i = -1$). Now if we take $\overline{x}$ to be the average 
value of $x_i$'s for those $l$ significant papers, then $E_r \sim 
\overline{x}(\frac{\overline{c}}{100\overline{n}})\sum_{i=1}^{l}s_i$.
With $N_c = \sum_{i=1}^{N_p}c_i \sim l \overline{c}$, we get the following,
$\Delta = \frac{100}{N_c}\times E_r
\sim \frac{100}{l\overline{c}}\times 
\overline{x}(\frac{\overline{c}}{100\overline{n}})\sum_{i=1}^{l}s_i$, or, 
$\Delta \sim \frac{\overline{x}}{l\overline{n}}\sum_{i=1}^{l}s_i$. We note 
that, if an individual's estimated contribution to a multi-author paper is 
more (less) 
than the average contribution of coauthors, then $s_i = + 1$ ($s_i = - 1$).
If all $s_i$'s are +1, then $\sum_{i=1}^{l}s_i = l$. On the other extreme,
if all $s_i$'s are -1, then $\sum_{i=1}^{l}s_i = -l$. In principle,
depending on the
details of the author's contributions made to the $l$ significant papers,
$\sum_{i=1}^{l}s_i$ can take any of the following possible values: $\{-l, -l+2, 
-l+4, \cdots, l\}$. Since the value of $\Delta$ can be different depending
on the value of $\sum_{i=1}^{l}s_i$, we will now calculate an expected 
value of $\Delta$ for an individual author. Noticing that a simple average
over all possible values of $\Delta$ is 0, we will here consider the 
root mean square value of $\Delta$ as its expected value. Once we know this 
root mean square value (denoted as $|\overline{\Delta}|$), we can 
say that, an individual's percentage share of credit for his/her works would 
be normally within ($I\pm|\overline{\Delta}|$)\% where the value of $I$ is 
given by Eq. \ref{iindex}. 
Now to calculate $|\overline{\Delta}|$, we first note that 
$s_i$'s are independent variables. This is because an author's amount of 
contribution to one paper does not presumably depend on his/her amount of 
contribution to another one. This independence of variables allows us to 
use some simple statistical results in estimating $|\overline{\Delta}|$. 
Now, these $l$ independent variables can take values in $2^l$ possible 
ways. For example, all the variables can be 1. This can happen in only one 
way ($^lC_0$) and in this case $\sum_{i=1}^{l}s_i = l$. Similarly, one
variable can be -1 and the rest can be 1. This can happen in $^lC_1$ ways and
in this case $\sum_{i=1}^{l}s_i = l-2$. In general $k$ variables can be -1 
and the rest ($l-k$) variables can be 1; this can happen in $^lC_k$ ways and 
here $\sum_{i=1}^{l}s_i = l-2k$. 
This counting helps us write the desired quantity in the following way: 
$|\overline{\Delta}| \sim \left(\frac{\overline{x}}{l\overline{n}}\right) 
\left(\frac{1}{2^l}\sum_{i=0}^l~^lC_i (l-2i)^2 \right)^{1/2}$. Some 
simple calculation shows that, 
$\left(\frac{1}{2^l}\sum_{i=0}^l~^lC_i (l-2i)^2 \right)^{1/2} = \sqrt{l}$.
Therefore, we get 
$|\overline{\Delta}| \sim \frac{\overline{x}}{\overline{n}\sqrt{l}}$.
Here we see that the value of 
$|\overline{\Delta}|$ gets smaller with increasing $l$ (and $\overline{n}$).
While a typical value of $|\overline{\Delta}|$ is expected to be less than 1, 
a typical value of $I$ is about 40. So here we conclude that the equidistribution 
of credit scheme gives us a reasonably good value of the $I$-index without much
statistical error.

\smallskip
{\it Argument} (2): 
It is possible to give a somewhat better mathematical argument, based on 
the {\it Central Limit Theorem} \cite{bhatt00} (CLT), to show that the 
value obtained from Eq. \ref{iindex} is the most probable value of the 
$I$-index (with an associated small standard deviation which decreases
with the increasing number of significant papers). A very careful 
analysis of the situation is needed here. As we discussed earlier, 
due to the inherent subjective nature of the analysis, it is hardly 
possible to decide who gets how much credit for a multi-author paper. 
If different experts independently evaluate the 
distribution of credit among the coauthors of a paper, then a particular 
author will get different values of credit from the different experts 
depending on how the evaluation was done. So the $I$-index for a 
researcher, defined in Eq. \ref{iidxdf}, will have different values when 
calculated by different experts. Which value 
shall we take? It would be recommended to take an average of these values. 
So what is the average or expectation value of the $I$-index if a large 
number of experts independently calculate it? Using the Central Limit 
Theorem we will now show that, within some reasonable 
assumptions, the average value of the $I$-index is what one gets by the 
scheme of equidistribution of credit (cf. Eq. \ref{iindex}). We are also 
interested in knowing the standard deviation about the average value, 
since a small deviation will allow us to confidently say that the 
average value is what an individual's share of credit is without much 
uncertainty. 

When a large number of experts independently evaluate the sharing of 
credit for a multi-author paper, the values of credit obtained by a 
particular author will follow some distribution. That is to say, an 
author will get a certain credit with some probability. 
Let for the $i$-th paper its $j$-th author gets $y_i^j$ credit with
the (marginal) probability density $K_i(y_i^j)$. In the joint probability
distribution of credits (for a particular paper $i$), the variables
$y_i^j$'s are not totally independent; they obey a singular 
constraint: $\sum_{j=1}^{n_i} y_i^j = c_i$ (with $0< y_i^j \le c_i$). 
So we see that the (random) variables $y_i^j$'s for
different $j$'s are not independent, even though $y_i^j$'s are totally
independent variables for different $i$'s (for an individual author $j$). 
This makes it easier to apply statistical theory to determine the 
probability distribution for the $I$-function defined for a specific 
author:
\begin{eqnarray}
\label{idstfn}
I(Y) = \frac{\sum_{i=1}^{N_p} y_i}{N_c}\times 100.
\end{eqnarray}
Note that the author index $j$ is dropped from the credit variables
$y$'s as we are focussing on a particular author. The symbol $Y$ denotes
the sum of all random variables ($y_i$'s). It may be 
noted that the $I$-function, defined for an individual, does not give 
a single value since each variable $y_i$ follows some distribution. 
The $I$-function gives a value with some probability; we are interested
in knowing the average value of the $I$-function and the standard 
deviation associated with it. 

Before we go further, let us briefly discuss what the CLT 
tells us. Let $X_1$, $X_2$, $\cdots$, $X_n$ are $n$ number of independent 
random variables with arbitrary distributions but each has a well-defined mean 
value ($E[X_i] = \mu_i$) and a well-defined variance ($var(X_i) = \sigma_i^2$). 
Now consider the function: $Y = \sum_{i=1}^n X_i$. The CLT assures us that, 
in the limit of large $n$, values of $Y$ will follow a {\it normal}
or {\it Gaussian} distribution with a mean given by 
$ E[Y] = \sum_{i=1}^n \mu_i$ and a variance given by 
$var(Y) = \sum_{i=1}^n \sigma_i^2$. This result from the CLT does not 
depend on the details of distributions of $X_i$'s, and is often valid even 
for a small $n$ \cite{note1}.

Since the variables $y_i$'s are essentially independent, in the limit of 
large number of papers ($N_p$), we can use the above statistical results 
to assure ourselves that the $I$-function will be a Gaussian in nature 
whose mean and variance can be given in terms of the means and variances 
of the variables $y_i$'s. To make things more quantitative, 
we now need to consider the means and the variances of $K_i$'s. 
Since the variable $y_i$ can take any value between 0 and $c_i$, and 
there are $n_i$ authors to share the total credit $c_i$, a reasonable 
assumption would be to take the mean value of the
variable $y_i$ to be $c_i/n_i$ (note: if we sum this over all coauthors
of $i$-th paper, we get back the total credit $c_i$). In fact, even if
the mean value of $y_i$ is not strictly $c_i/n_i$, we will still normally
have the same results that follow. Argument for this will be given
soon after I write down the mean and variance of the $I$-function.
Since the range of the variable $y_i$ is finite, its variance will also 
be finite (for any regular distribution); let us for the time 
being consider  $\sigma_i^2$ ($<\infty$) be its variance.
If we now use the CLT results for  
$I(Y)$, we get the following: in the limit of large $N_p$, the
values of $I(Y)$ will be distributed in a {\it normal} or 
{\it Gaussian} distribution with the mean
$\frac{100}{N_c}\sum_{i=1}^{N_p} c_i/n_i$ (i.e. the $I$-index
defined in Eq. \ref{iindex}) and the variance
$\Sigma^2 = \frac{100^2}{N_c^2}\sum_{i=1}^{N_p}\sigma^2_i$. 
It may be noted that here we have used following two general 
relations: $E[a X_i] = a E[X_i]$ and $var(a X_i) = a^2 ~var(X_i)$, 
where $a$ is any constant.

Now I will argue that even if the mean of $y_i$ is not strictly $c_i/n_i$,
we will still normally have the same mean for the $I$-function.
The reasoning goes exactly like the {\it Argument} (1) 
given before. 
Statistically, for some variables corresponding mean can be more 
than $c_i/n_i$ (i.e., $E[y_i] \ge c_i/n_i$) 
and for others the mean can be less than that 
(i.e., $E[y_i] < c_i/n_i$). Therefore when we calculate 
the sum of the means of $y_i$'s, we expect that the result will not be 
much different than $\sum_{i=1}^{N_p} c_i/n_i$.
Now whatever (small) difference it might have, that
will be further weakened by the large denominator factor $N_c$ present
in the definition of the $I$-function. So here we conclude that, in
all normal cases, the mean value of the $I$-function is
$\frac{100}{N_c}\sum_{i=1}^{N_p} c_i/n_i$ without much significant 
deviation.

Now we will analyze whether the $I$-function has broad or narrow peak 
about its mean value.
A narrow peak about the mean value will allow us to confidently
say that, an author's $I$-index is what one gets from
Eq. \ref{iindex}.

For the distribution of $y_i$, the standard deviation $\sigma_i$ is expected
to depend on $c_i$ (this is because, normally larger is the range of
a variable, wider is the distribution; here the variable $y_i$ varies
from 0 to $c_i$). We assume $\sigma_i$ to be some percentage of the mean
value $c_i/n_i$ of the distribution, i.e.,
$\sigma_i \sim  c_i/n_i \times x_i/100$ ($x_i$ takes values between,
say, 0 and 20). Let us now consider that an author has $l$ number of
significant papers so that the total number of citations for these $l$
papers is much larger than the total number of citations for the rest of
the papers (i.e., $\sum_{i=1}^lc_i\gg\sum_{i=l+1}^{N_p}c_i$ 
when papers are arranged in the descending order of citation count). 
The value of $l$ can be assumed to be the $h$-index of the
author. If $\overline{c}$ and $\overline{n}$ are respectively the average
number of citations and the average number of authors for those $l$
significant papers, then $N_c = \sum_{i=1}^{N_p}c_i 
\sim l\overline{c}$ and $\sum_{i=1}^{N_p}\sigma^2_i 
\sim \sum_{i=1}^{N_p} (\frac{c_i}{n_i}\frac{x_i}{100})^2 \sim 
l \frac{\overline{c}^2}{\overline{n}^2}\frac{\overline{x}^2}{100^2}$, 
where $\overline{x}$ is the average value of $x_i$ for those $l$ 
significant papers. This implies that, 
$\Sigma^2 = \frac{100^2}{N_c^2}\sum_{i=1}^{N_p}\sigma^2_i \sim 
\frac{100^2}{l^2 \overline{c}^2} \times 
\frac{l \overline{c}^2 \overline{x}^2}{\overline{n}^2 100^2}$, or
$\Sigma \sim \frac{\overline{x}}{\overline{n}\sqrt{l}}$. 
We see that the value of $\Sigma$ gets smaller with an increase in the 
values of $l$ (or $h$-index) and $\overline{n}$.
A typical value of the standard deviation $\Sigma$ is expected to be 
less than 1 whereas a typical value of the mean value 
of the $I$-function is about 40. 
So we conclude that, normally the $I$-function defined in 
Eq. \ref{idstfn} has a very sharp Gaussian distribution
about its mean value given by the $I$-index (cf. Eq. \ref{iindex}). This
allows us to say that {\it the most probable value of the $I$-index can be
obtained by a simple scheme of equidistribution of credit among the coauthors
of a paper. Uncertainty (statistical standard deviation) associated with the
value is normally very small (especially for authors with high $h$-index)}.

\smallskip
In the following I will now discuss some of the main features/characteristics 
of the $I$-index.

\smallskip
{\it Characteristic} (a):
Unlike the $h$-index or $N_c$ (= $\sum_{i=1}^{N_p}c_i$), the $I$-index
is expected to be a very slowly varying function of time. 
The $h$-index is linear in time while $N_c$ is quadratic
in time \cite{hirsch05}. Similar to $N_c$, $C_{share} = \sum_{i=1}^{N_p}c_i/n_i$ 
is also expected to be quadratic in time since both $N_c$ and $C_{share}$ are 
essentially linear sum of $c_i$'s (see argument given in 
Ref. \cite{hirsch05}). 
Now we assume that, $N_c = a_1 t+a_2 t^2$ and $C_{share} = b_1 t+b_2 t^2$, 
where $t$ is the career span of a scientist (see Sec. \ref{sec:3.1}), and 
$a_1$, $a_2$, $b_1$ and $b_2$ are some constants (author dependent).
This leads us to $I$ as a following function of time, 
$I = 100\times\frac{C_{share}}{N_c} = 
100\times\frac{b_1 t+b_2 t^2}{a_1 t+a_2 t^2}$, or 
$I = 100\times\frac{b_2 + b_1/t}{a_2 +a_1/t}$. It is now easy to see why the 
$I$-index is expected to be a very slowly varying function of time.

For a similar reason, the $I$-index will not be much affected by career break or
low activity (due to some severe medical condition,
family tragedy or importantly a female researcher's motherhood). 
We note that a career break/low activity would affect both $N_c$
and $C_{share}$ in a similar way. So their ratio i.e. the $I$-index is expected 
to be mostly free of the effects caused by these important subjective issues.

\smallskip
{\it Characteristic} (b):
 Normally an author's affiliation, seniority or popularity affects 
the citations ($c_i$'s) received by his/her papers. As a result 
both $C_{share}=\sum_{i=1}^{N_p} c_i/n_i$ and $N_c = \sum_{i=1}^{N_p} c_i$ would 
depend on those factors. Since both the quantities, $C_{share}$ and $N_c$, 
are linear functions of $c_i$'s, we expect that both of them will be influenced 
in a similar way by those factors. Now as the
$I$-index is defined as the ratio between those two quantities, it is expected
that those subjective issues will not help better one's $I$-index. 
The essential functional difference between the $I$-index and 
$N_c$ or $h$-index is that, unlike the later two, the $I$-index is a relative
quantity which effectively quantifies what fraction of the total credit an 
individual entitled to get for his/her papers. Being a relative quantity, we expect
the $I$-index to be mostly independent of all the subjective issues mentioned.

\smallskip
For the properties of the $I$-index stated above (cf. (a) and (b)), it will not
be unfair to compare
values of this index for the authors with different seniorities or
affiliations/popularities. 

\smallskip
{\it Characteristic} (c) 
The $I$-index can only be improved if
a researcher starts publishing single-author or a-few-author impactful papers. Here
it may be noted that even if someone manages to improve his/her $h$-index and $N_c$
by doing large number of collaborations, the $I$-index may not increase in this way,
and sometimes it may decrease!
Unlike $N_c$ or the $h$-index, the $I$-index is not a
monotonically increasing function of time. For example, its value may decrease if
a paper with a large number of authors starts getting highly cited or a researcher
starts publishing large number of highly collaborative works.

\smallskip
{\it Characteristic} (d): 
Unlike $N_c$ and $h$-index, the $I$-index is a bounded parameter. We see 
from Eq. \ref{iindex} that, if $n_i =1$ for all $i$, then $I = 100\%$, and if
$n_i$'s are very large, then $I$ will be very small. For any author this index 
takes a value between 0 and 100. Theoretically, $0\% < I \le 100 \%$ for any 
fixed non-zero values of $N_c$ and $h$-index. 

\section{The triplet: $N_c$, $h$-index, $I$-index}
\label{sec:3}
In this section we will see how $N_c$, $h$-index and $I$-index are three 
independent parameters and together they can provide us a comprehensive idea 
of an author's overall research merit. We will also see advantages of choosing 
them over other available parameters.

First we note that, irrespective of the values of $N_c$ and $h$-index, 
the $I$-index can take any possible value between 0 and 100 depending on the number 
of coauthors of the published papers (as explained above, see 
{\it Characteristic} (d) of the $I$-index). 
Theoretically, $0\% < I \le 100 \%$ for any fixed non-zero values
of $N_c$ and $h$-index. 

Now for a fixed non-zero value of the $h$-index, the minimum possible value of
$N_c$ is $h^2$ while the maximum value can be any large number 
depending on the number of citations received by the individual 
papers within the $h$-core. Theoretically,
$h^2 \le N_c < \infty$ for any fixed non-zero values of the $h$-index and $I$-index.

For a fixed non-zero value of $N_c$, the minimum possible value of the $h$-index 
is 1, while the maximum value is $\lfloor \sqrt{N_c} \rfloor$ if the number of 
papers $N_p \ge \lfloor \sqrt{N_c} \rfloor$ else the maximum value is $N_p$. 
Theoretically, $1 \le h \le h_{max}$ for any fixed non-zero values of $N_c$ and 
$I$-index. Here $h_{max} = \lfloor \sqrt{N_c} \rfloor$ when 
$N_p \ge \lfloor \sqrt{N_c} \rfloor$,
otherwise $h_{max} = N_p$. It may be noted here that 
$\lfloor x \rfloor$ is the usual mathematical floor function.

I will now give three {\it elementary} examples to illustrate that 
the three parameters are independent and that each parameter gives an important 
information which is not contained in other two parameters. First let us consider
two researchers each with 10 papers, and for both of them, their papers are cited  
followingly (when arranged in the descending order of citation count): 
first paper is cited 10 
times, the second one is cited 9 times, and so on (i.e., the $i$-th paper is cited 
($11-i$) times). In this example, $N_c = 55$ and $h = 5$ for both the authors. 
In addition, if we now consider that the first researcher wrote all his/her 
papers with one more author (total two authors per paper) and the second 
researcher wrote all his/her papers with two more authors (total three authors 
per paper), then $I = 50\%$ for the first researcher and $I = 33.33\%$ for the 
second researcher. 
This shows that, even when two researchers have the same $N_c$ and $h$ values, they 
can have quite different $I$ values depending on the number of coauthors. 
A smaller value of $I$ signifies that the researcher do more collaborative work. 
In the second example, consider that the first 
researcher has 12 papers, each cited 8 times and coauthored by two while the second
researcher has 10 papers, each cited 8 times and coauthored by two. In this case, 
$h = 8$ and $I = 50\%$ for both the researchers but $N_c = 96$ for the first 
researcher while $N_c = 80$ for the second researcher. So here, the total scientific 
impact of the first researcher is more than that of the other researcher even though 
their $h$ and $I$ values are same. 
In the third example, 
consider that the first researcher has 10 papers, each cited 8 times and coauthored 
by two while the second researcher has 20 papers, each cited 4 times and coauthored 
by two. In this case, $N_c = 80$ and $I = 50\%$ for both the researchers but $h = 8$ 
for the first researcher and $h = 4$ for the other researcher. In this example, 
the first researcher has more significant papers than the other researcher, or in 
other words, the first researcher's quality of research work is better than that of 
the second researcher even though their $N_C$ and $I$ values are same.  

From the above discussions it is clear that $N_c$, $h$-index and the $I$-index can 
take values independently (within their bounds).
These three parameters or metrics quantify three
most important aspects of a researcher's scholarly output -quantity, quality and a
researcher's own role in his/her overall success. Each of these independent parameters
carries important new informations; if we miss one, the description of a researcher's
merit will be highly incomplete. This shows why a single parameter, however smartly
defined, would be insufficient and gross in describing a researcher's scholarly output.

Now I will discuss, instead of other possible parameters, why I choose 
$N_c$, $h$-index and $I$-index as the preferred ones to quantify the three separate  
aspects of an author's research output.

The $h$-index is known to be the best single parameter which somewhat successfully
quantifies the first two aspects of one's research output, i.e., the collective impact 
and the productivity (or in other way, the quality and quantity). 
But most of the time this parameter highly under-estimates
the total impact of an author's research output. For example, two authors 
having same value of $h$-index can have widely different collective impact if one of the 
authors has some very highly cited papers within his/her $h$-core. 
This necessitate us to choose a separate 
parameter to represent the collective impact of an author's research output; the total 
citations or $N_c$ is the natural choice for this purpose. The advantages
of using these two parameters are that they have simple and easy-to-calculate definitions
and can provide very efficient and comprehensive description of the first two aspects of 
one's research output.

The concern for accounting coauthors' contributions is not new and has been considered
in many previous works
\cite{hirsch10,pepe12,kurtz05,schreiber09,batista06,egghe08,pal15,ausloos15,%
biswal13,galam11,liu12}.
Now I will argue why the $I$-index does a reasonably good job in 
quantifying the third aspect of one's research output, i.e., an author's own contribution
in his/her published works. In the most of the related works I know, all three aspects 
of one's research output were tried to be quantified by a single unbound parameter or by
a coauthor ranking algorithm. But as we have emphasized several times, 
any single parameter (or any
ranking algorithm which assigns a score to each coauthor) will be unsatisfactory in
describing an author's research output due to serious loss of informations. 
Moreover, it is not clear from those works 
whether the consideration of coauthorship would discourage true collaborations (for 
notable exception, see Ref. \cite{hirsch10}). No bibliometric indicator should 
discourage scientists from doing honest collaboration which is imperative for the 
progress and betterment of science. The ranking algorithms have additional problems.
Generally they are
computationally extensive for large number of authors sharing even larger number of
papers. In practice hundreds of authors can be connected to each other by 
a coauthorship network and they may share thousands of papers  
(sometimes it is not even practical to get a complete set of authors sharing papers
among them). Since in principle the ranking algorithms should 
simultaneously rank all these authors (and also papers) by solving equation of large 
matrices (representing authors, papers and their inter connections), it looks very unlikely
that these algorithms can practically resolve the coauthorship issue. 
Additionally, due to complex computation (normally involves iterative matrix
manipulations \cite{pal15}), the ranking looses intuitive meaning (or comprehensiveness) 
for the wider population. 
In contrast to these works, in this paper we do not try to quantify all the 
aspects of one's research output by a single parameter. The $I$-index proposed here is a 
complimentary metric, meant to quantify only one aspect of an individual's research output. 
It has a simple intuitive meaning (cf. Eq. \ref{iidxdf}), 
is easy to calculate and argued to provide a reasonably good measure 
even with a simple scheme of equidistribution of credit (see {\it Argument} (1) and 
{\it Argument} (2) given in Sec. \ref{sec:2}). The $I$-index, being a bounded 
parameter (varies from 0 to 100), will be very helpful in judging authors according to 
their performance in the third aspect of research output. A high value of the $I$-index 
signifies that the author works more independently (see also discussions in 
Sec. \ref{sec:3.1} and Sec. \ref{sec:4}).

An important advantage of separately considering $I$-index besides
$N_c$ and the $h$-index is that, it will discourage the unethical practice of giving/taking 
authorship to/by non-contribution authors. This will not probably though deter
scientists from doing true collaborations, as otherwise their $N_c$ and the $h$-index
will not improve (see also Sec. \ref{sec:3.1}).

\subsection{Ranking of authors}
\label{sec:3.1}
It is always difficult to make a merit list for authors. But when it is needed, how do we 
do it? Here I will discuss some practical ways of ranking authors. 

First I will discuss how this can be done using the three independent parameters 
deliberated in this paper. 
In fact using three independent parameters the ranking can be done in different ways 
depending on which aspect of research is considered to be more important (for, say, 
a particular job).
Three independent parameters naturally gives more freedom to the employers to choose 
candidates of their requirements. For example, considering $h$-index is the most important 
parameter among the three parameters, first one can try to rank authors according to their 
$h$ values. Surely there will be many authors with same (or close) $h$ values. One of the 
reasons for the occurrence of degeneracy is that the $h$-index takes only discrete integer 
values. The authors with same or close $h$ values can be ranked using the $I$-index. An 
author with better $I$ value should rank higher. In the next step, if these two parameters 
does not help to resolve the ranking issue among a group of researchers then their $N_c$ 
values can be used to see who is the better performer.
If for a particular job employers are looking for a researcher who can work independently, 
then they probably can give more importance to the $I$-index. In this case, among the 
researchers with their $h$ values within a fixed range, the employers can choose the 
person who has highest $I$ value.
 
As I already emphasized, a single parameter/metric will not be sufficient and reliable 
in describing an author's research merit. With this fact in mind, we now ask, which  
parameter shall we use if for some practical reasons it is needed to rank authors 
by a single parameter? For this purpose I will now define a normalized $h$-index 
(written as $\widetilde{h}$-index) 
which combines the effects/impacts of both $h$-index and $I$-index in a rational 
way. Subsequently I also propose $\widetilde{h}_T$-index which 
additionally takes care of the seniority issue.\\

\noindent
{\underline{\it $\widetilde{h}$-index and $\widetilde{h}_T$-index}}:
Here idea is to estimate how much an author would have achieved if he/she had worked 
alone. Roughly an author will have $N_a = N_c * I/100$ citations for his/her works 
if he/she worked alone (see definition of $I$-index, Eq. \ref{iindex}). 
It is shown in Ref. \cite{hirsch05}
that the total number of citation ($N_c$) is proportional to $h^2$ (this is a
general trend with the proportionality constant varies for different authors). 
Therefore, $N_a = g_1 h^2 * I/100$; where $g_1$ is the  proportionality constant. 
Now if $\widetilde{h}$ is the expected $h$-index of the author if he/she had worked 
alone, then $N_a$ should be proportional to $\widetilde{h}^2$, i.e., 
$N_a = g_2 \widetilde{h}^2$ with $g_2$ being another proportionality constant. 
Comparing two expressions of $N_a$, we get the following relation: 
$\widetilde{h} = (\sqrt{\frac{g_1}{g_2}})~h*\sqrt{I}/10$. It is not easy to 
find any simple relation between the two constants $g_1$ and $g_2$. Here I present 
a rough argument to show that, for a given individual, the values of these two 
constants would not be much different. According to the simplest possible model 
discussed in Ref. \cite{hirsch05}, $g_1 = \frac{(1+c/p)^2}{2c/p}$, where the 
researcher publishes $p$ papers per year and each published paper gets $c$ 
new citations per year in every subsequent year. Now if the researcher had worked
alone, the value of $p$ would have been smaller. Since an effective collaboration 
enhances quality of papers, we can expect that $c$ would also get smaller if the 
researcher works alone. Due to the collective or cooperative effect of 
collaboration, the sum of impacts of independent individuals is expected to be 
smaller than the total impact of the works done in collaboration by those individuals.
Going by this argument, we can say that the ratio $c/p$ will not be much different 
depending on whether a researcher works alone or in collaborations. This implies 
that, the value of $g_2$ is expected to be reasonably close to $g_1$. 
This is in accordance with the fact 
that, irrespective of the collaboration details of researchers, the proportionality 
constant $g_1$ takes values from a small range of numbers 
(between 3 and 5 \cite{hirsch05}). Now since the square root of a positive number is 
always closer to 1 than the number itself ($|1-\sqrt{x}|\le|1-x|$ with $x>0$),  
we expect that $\sqrt{\frac{g_1}{g_2}}$ will be very close to 1 even though 
$\frac{g_1}{g_2}$ is somewhat away from 1. Now taking 
$\sqrt{\frac{g_1}{g_2}} \approx 1$, we get the following formula for 
the normalized value of the $h$-index, 
\begin{eqnarray}
\widetilde{h} = h*\sqrt{I}/10.
\label{hnrm}
\end{eqnarray}
We note that, if an author publishes only single-author papers,
then his/her $I = 100$, and consequently his/her $\widetilde{h} = h$. This is in 
accordance with what one expects for a researcher who always works alone. The 
experimentalists do more collaborative works than the theorists; so compared to a 
theorist, an experimentalist will normally have higher value of $h$ and 
lower value of $I$. 
This trend can be seen in the next subsection on results (see Table \ref{tbl1}). 
For a theorist and an experimentalist of presumably same calibre, their values of 
$\widetilde{h}$-index should be very close even though their $h$ and $I$ values 
are quite different. Interestingly this is what we observe in our analysis of 
some established authors (see Sec. \ref{sec:3.2}). 

Since $\widetilde{h}$ depends on both $h$ and $I$, to improve
the value of $\widetilde{h}$-index, a researcher needs to better both those parameters or 
at least better one parameter keeping another relatively fixed. 
Advantage of considering the 
$\widetilde{h}$-index over the original $h$-index is that, it will discourage researchers 
to involve in unethical practice of giving/taking authorship without substantial
contribution. If they do, their $I$-index will reduce and as a consequence their 
$\widetilde{h}$-index will also be badly affected. But probably this will not dissuade
researchers to do true collaboration, as otherwise their $h$-index will not improve much 
and as a result $\widetilde{h}$-index will not get better.

It should be noted that the $\widetilde{h}$-index is not an independent parameter,
it is a derived parameter/metric proposed here to help rank authors using a single 
parameter. This parameter does not take into consideration the issue of seniority or 
length of research career. This can be done by dividing $\widetilde{h}$ by the length 
of an author's research career. If $T$ is the time (in years) between the first 
publication (at least once cited) and the last published one, then we define,
\begin{eqnarray}
\widetilde{h}_T = \widetilde{h}/T.
\label{hnrm_t}
\end{eqnarray} 
This parameter ($\widetilde{h}_T$) takes into consideration both the issues of 
coauthorship and the length of research career. Though this simple division by 
career length has some problems. It will be unfavorable for the authors  who had taken 
career breaks. At the same time it will favor the authors whose careers have ended. 
This second problem can be somewhat circumvented by taking $T$ as the time between 
the first publication and 
the time of data collection. Here it may be noted that, in mathematical sense, 
$\widetilde{h}_T$ is not a derived parameter since $T$ is an independent parameter.

\begin{table*}[tp]
\caption{The values of the parameters/metrics $N_c$, $h$, $I$, $\widetilde{h}$ and 
$\widetilde{h}_T$ are given for some established
authors. Age of an author is given within bracket just after his/her name. Under
each author's name his/her specialization and major awards (if any) are given.
Here,
TP = Theoretical Physics, EP = Experimental Physics, HE = High Energy physics,
CM = Condensed Matter physics, AMO = Atomic, Molecular and Optical physics,
QI = Quantum Information science, FM = Field Medalist, NL = Nobel Laureate.}
\begin{center}
\begin{tabular}{l|c|c|c||c|c}
\hline
 & & & &\\  %\hline
Author & $N_c$ & $h$-index & $I$-index (\%) & 
$\widetilde{h} = h*\sqrt{I}/10$ & 
$\widetilde{h}_T=\widetilde{h}/T$\\
 & & & &\\  \hline
E. Witten (63) &  &  &  & &\\
(TP-HE, FM) & 166563 & 179 & 74.35 & 154.3& 3.9\\
 & & & & & \\  \hline
A. Sen (59) & & & & & \\
(TP-HE) & 25967 & 85 & 81.62 & 76.8 & 2.3\\
 & & & & & \\  \hline
C.W.J. Beenakker (55) & & & & & \\
(TP-CM) & 29983 & 83 & 50.12 & 58.8 & 1.8\\
 & & & & & \\  \hline
D.J. Gross (74) & & & & & \\
(TP-HE, NL) & 44292 & 83 & 45.64 & 56.1 & 1.1\\
 & & & & & \\  \hline
T.W. H\"{a}nsch (73) & & & & & \\
(EP-AMO, NL) & 51719 & 107 & 23.97 & 52.4 & 1.1\\
 & & & & & \\  \hline
C.L. Kane (52) & & & & & \\
(TP-CM) & 29471 & 55 & 43.26 & 36.2 & 1.3\\
 & & & & & \\  \hline
A.E. Nelson (57) & & & & &\\
(TP-HE) & 17153 & 52 & 37.81 & 32.0 & 0.9\\
 & & & & & \\  \hline
C. Monroe (49) & & & & & \\
(EP-AMO-QI) & 24774 & 60 & 19.85 & 26.7 & 1.0\\
 & & & & & \\  \hline
\end{tabular}
\end{center}
\label{tbl1}
\end{table*}

\subsection{Some results}
\label{sec:3.2}
I have estimated three independent parameters/metrics ($N_c$, $h$ and $I$)
for some of the established researchers. List is prepared carefully to 
represent researchers working in different fields and belonging to different age groups 
(there is 25 years of age gap between youngest and oldest researcher). 
The results can be found in Table \ref{tbl1}. In the last two columns of the table
values of the other two parameter ($\widetilde{h}$-index and $\widetilde{h}_T$-index)
are also given. As I discussed in Sec. \ref{sec:3.1}, ranking can be 
done in different ways depending on how we analyze the research output. In addition, 
since the listed researchers work in different (sub)fields, it may not be 
appropriate to compare their performance without considering the publication/citation 
trends in the (sub)fields (for a discussion, 
see \cite{pepe12}). In any case, for the completeness of our analysis in this paper, 
they are ranked in the table according to their $\widetilde{h}$ values. 
We may here note that, generally 
those having high $\widetilde{h}$-index, have high $\widetilde{h}_T$-index. For two 
authors with close $\widetilde{h}$ value, one may have lower $\widetilde{h}_T$ value
than the other if he/she takes a career break for some reason. This is because, a 
career break acts more harsh on $\widetilde{h}_T$ than $\widetilde{h}$.    
We also see from the table that the experimentalists have lower value of $I$-index 
than the theorists. This is because experimentalists generally do more collaborations 
than theorists (an experimental paper normally has more authors than a theory paper). 
For the same reason, generally the experimentalists have higher $h$-index than 
the theorists of their age group. This discipline dependency of these two parameters
is the reason we choose $\widetilde{h}$-index to decide the ranking in the table 
($\widetilde{h}$-index combines the effects/impacts of both $h$-index and $I$-index 
in a rational way).
It is here interesting to note that, for the two Noble Laureates (D.J. Gross, a 
theorist and T.W. H\"{a}nsch, an experimentalist), the research output measured 
by $\widetilde{h}$ or $\widetilde{h}_T$ is same or very close even though 
their $h$-index and $I$-index are quite different.  

The parameters in the table 
are extracted from the data collected manually in July, 2015
from {\it Google Scholar Citation}. In the calculation of parameters, not only the
original research papers, other scholarly works like review articles and books are 
also considered. Some practical issues may appear while estimating these parameters.
For example: different chapters of a book can be written by different authors. In this
case if the total citations of the book is available, then that citation number can be
first divided by the number of chapters and next this credit per chapter can be divided
among the coauthors of a chapter to determine how much credit one author should get.
If the detail author information of a scholarly work is missing, then the $I$-index
should be calculated simply ignoring that particular work.

\section{Conclusion}% and outlook}
\label{sec:4}
In this paper I have tried to establish a rational and objective  
framework for analyzing scientists' research outputs. Three most important aspects of 
someone's research performance have been identified -collective impact, productivity 
and author's own contribution in his/her published works. It is emphasized that we 
need three independent parameters/metrics to quantify those three separate aspects
reliably. A single parameter will be insufficient and gross in describing an 
author's research performance due to serious loss of informations. 
A practical advantage of using three independent parameters for analysis is that it 
will give employers more freedom to choose candidates according to their requirement. 
I have suggested following three parameters for the purpose: the total number
of citations ($N_c$), the $h$-index and the newly defined $I$-index. The $I$-index
is defined as an author's claim for the percentage of total citations received by 
his/her papers. Besides its simple and comprehensible meaning, this index is very 
easy to calculate and argued to be almost independent of most of the subjective issues 
like affiliation, seniority or career break. It is also argued using the 
{\it central limit theorem} that,
the most probable value of the $I$-index can be obtained by the simple scheme of 
equidistribution of credit among the coauthors of a paper. Uncertainty associated with 
the value is normally very small.

It will be highly unfair for researchers working alone or in small groups if we consider 
only $N_c$ and $h$-index to judge their performance. The researchers sharing time with 
many collaborators will normally have large number of papers and
consequently have higher $N_c$ and $h$-index. So it is crucial to distribute credit among
the coauthors and measure how much contribution one has in his/her scientific achievement.
The new index (i.e., $I$-index) 
proposed in this paper tries to address this crucial issue.
A larger value of the $I$-index signifies that the author works more independently 
(this is why the $I$-index can be considered as the {\it Independence}-index). 
A practical advantage of considering this $I$-index along with $N_c$ and the $h$-index 
is that, it will discourage scientists from engaging in the unethical practice 
of giving/taking authorships to/by non-contributing scientists. This will, though, 
probably not deter scientists from doing true collaborations, as otherwise
their $N_c$ and the $h$-index will not improve.

In this work we have also defined $\widetilde{h}$-index, and subsequently $\widetilde{h}_T$-index, 
to rank authors if for some practical reasons it is needed to rank them using a single parameter. 
Unlike the $h$-index, the $\widetilde{h}$-index takes into consideration the crucial issue of
coauthors' contributions, while $\widetilde{h}_T$-index additionally takes care of the seniority 
issue.

Since low value of the $I$-index signifies a more collaborative nature 
of one's work, we can define a {\it Collaboration}-index or $C$-index, as a complementary index 
to the $I$-index: $C = 100 - I$. Note that, like $I$, $C$ also takes values between 0 and 100. 
A larger $C$  value for a researcher indicates that his/her work is more collaborative in nature.
In future study, the average $C$-index for the scientists working in a 
particular field or in a particular institute can be estimated; this will tell us in which 
field or institute scientists do more collaborative works than others. Similarly the average 
values of the $C$-index for different countries can be calculated to see in which country 
scientists do more collaborative work.

\begin{acknowledgements}
The author thanks CEFIPRA for financial support.
\end{acknowledgements}

% BibTeX users please use one of
%\bibliographystyle{spbasic}      % basic style, author-year citations
%\bibliographystyle{spmpsci}      % mathematics and physical sciences
%\bibliographystyle{spphys}       % APS-like style for physics
%\bibliography{}   % name your BibTeX data base

\begin{thebibliography}{}
\bibitem{hirsch05}Hirsch, J. E. (2005). An index to quantify an individual’s 
scientific research output. {\it PNAS}, {\it 46}, 16569.
\bibitem{hirsch10}Hirsch, J. E. (2010). An index to quantify an individual’s 
scientific research output that takes into account the effect of multiple 
coauthorship. {\it Scientometrics}, {\it 85}, 741.
\bibitem{pepe12}Pepe, A., \& Kurtz, M. J. (2012). A Measure of Total Research Impact 
Independent of Time and Discipline. {\it PLoS ONE}, {\it 7}, e46428.
\bibitem{kurtz05} Kurtz, M. J., Eichhorn, G., Accomazzi, A., Grant, C., Demleitner, M.,
Murray, S. S., Martimbeau, N., \& Elwell, B. (2005). The Bibliometric Properties
of Article Readership Information. {\it J. Am. Soc. Inf. Sci. Techn.}, {\it 56}, 111.
\bibitem{bhatt00} Bhattacharya, R., \& Waymire, E. C. (2000). {\it A Basic Course
in Probability Theory}. USA: Springer.
\bibitem{note1}A brief note: the independence of the variables ($X_i$'s) itself 
implies that, for any $n\ge2$, $E[Y] = \sum_{i=1}^n E[X_i]$ and 
$var(Y) = \sum_{i=1}^n var(X_i)$ (Bienaym\'{e} formula). This though does not say 
anything about the shape/form of the distribution of $Y$. We employ the CLT because it
provides additional information that the distribution of $Y$ will be a Gaussian
in the large $n$ limit.
\bibitem{schreiber09}Schreiber, M. (2009). A case study of the modified Hirsch
index $h_m$ accounting for multiple coauthors. 
{\it J. Am. Soc. Inf. Sci. Techn.}, {\it 60}, 1274.
\bibitem{batista06}Batista, P. D., Campiteli, M. G., Kinouchi, O., \& Martinez, A. S.
(2006). Is it possible to compare researchers with different scientific interests?
{\it Scientometrics}, {\it 68}, 179.
\bibitem{egghe08}Egghe, L. (2008). Mathematical theory of the h- and g-index in case 
of fractional counting of authorship. {\it J. Am. Soc. Inf. Sci. Techn.}, {\it 59}, 1608.
\bibitem{pal15}Pal, A., \&  Ruj, S. (2015). CITEX: A new citation index to measure the 
relative importance of authors and papers in scientific publications. {\it 2015 IEEE 
International Conference on Communications (ICC)}. pp. 1256-1261. 
\bibitem{ausloos15}Ausloos, M. (2015). Assessing the true role of coauthors in the 
h-index measure of an author scientific impact. {\it Physica A}, {\it 422} 136.
\bibitem{biswal13}Biswal, A. K. (2013). An Absolute Index (Ab-index) to Measure a 
Researcher’s Useful Contributions and Productivity. {\it PLoS ONE}, {\it 8}, e84334.
\bibitem{galam11}Galam, S. (2011). Tailor based allocations for multiple authorship:
a fractional gh-index. {\it Scientometrics}, {\it 89}, 365. 
\bibitem{liu12}Liu, X. Z., \& Fang, H. (2012). Modifying h-index by allocating credit 
of multi-authored papers whose author names rank based on contribution. {\it Journal of 
Informetrics}, {\it 6}, 557.
\end{thebibliography}

% Non-BibTeX users please use

\end{document}